\documentclass[3p,times,procedia]{elsarticle}
\usepackage{ecrc}
\usepackage{listings}

\usepackage{dirtytalk}
\usepackage{caption}

\usepackage[colorinlistoftodos,prependcaption]{todonotes}

\usepackage[symbol]{footmisc}

\volume{00}

\firstpage{1}

\journalname{Procedia Computer Science}

\runauth{Umutcan \c{S}im\c{s}ek et al.}

\jid{procs}

\usepackage{amssymb}

\usepackage[figuresright]{rotating}

\usepackage{multirow}

\begin{document}

\begin{frontmatter}



\dochead{13th International Conference on Semantic Systems}

\title{Machine Readable Web APIs with Schema.org Action Annotations\footnote[2]{Throughout the paper, the term "Web API" refers to all HTTP APIs that adopt a certain resource hierarchy and run over HTTP. For the sake of clarity, we ignore the fact that not all "RESTful APIs" are following all REST architectural design principles. Throughout the paper "RESTful Web Service/API and "Web API" terms may be used interchangeably}}



\author{Umutcan \c{S}im\c{s}ek\corref{cor1}}
\author{Elias K\"{a}rle}
\author	{Dieter Fensel}

\address{STI Innsbruck, Department of Computer Science, University of Innsbruck, Technikerstrasse 21a 6020, Innsbruck Austria}

\begin{abstract}
The schema.org initiative led by the four major search engines curates a vocabulary for describing web content. The number of semantic annotations on the web are increasing, mostly due to the industrial incentives provided by those search engines. The annotations are not only consumed by search engines, but also by other automated agents like intelligent personal assistants (IPAs). However, only annotating data is not enough for automated agents to reach their full potential. Web APIs should be also annotated for automating service consumption, so the IPAs can complete tasks like booking a hotel room or buying a ticket for an event on the fly. Although there has been a vast amount of effort in the semantic web services field, the approaches did not gain too much adoption outside of academia, mainly due to lack of concrete incentives and steep learning curves. In this paper, we suggest a lightweight, bottom-up approach based on schema.org actions to annotate Web APIs. We analyse schema.org vocabulary in the scope of lightweight semantic web services literature and propose extensions where necessary. We show that schema.org actions could be a suitable vocabulary for Web API description. We demonstrate our work by annotating existing Web APIs of accommodation service providers. Additionally, we briefly demonstrate how these APIs can be used dynamically, for example, by a dialogue system.
\end{abstract}

\cortext[cor1]{Corresponding author}
\end{frontmatter}

\email{umutcan.simsek@sti2.at}


\vspace*{-5pt}


\section{Introduction}
\label{sec:introduction}

The semantic annotation of web content is realizing the vision of making the web machine readable. More than a decade ago, researchers have identified that the challenge is not only bringing semantics to the data on the web, but also to web services, in order to enable automated agents to understand them and automate web service tasks \cite{Burstein:2002:DWS:646996.711291} \cite{Fensel2002}. Initial efforts have been mostly focused on SOAP services. The semantic description of RESTful web services came later as they gained significant popularity within the last decade due to their lightweight approach and flexibility, for instance in terms of supported data formats. 

However, outside of academia, the adoption of semantic web services have remained quite low. The main reason for the weak adoption is a so called "chicken-egg problem", that is, there is no interest in application development since there are no annotated web services and there is no annotation effort since there are no applications. The aforementioned approaches are semantically very strong and well designed, however, for the web service providers they look challenging \cite{Lanthaler2011}. We aim to overcome this chicken-egg problem by lowering the entry barrier to the semantically annotated Web APIs. For that, we utilize the already well adopted schema.org vocabulary and try to create a lightweight semantic web services vocabulary based on schema.org actions. This way, the machine readable Web APIs can be consumed by agents like goal-oriented dialogue systems (e.g. Intelligent Personal Assistants) to complete tasks like purchasing products or booking hotel rooms on the fly.

In this paper, we propose an approach for using schema.org actions as a lightweight Web API description vocabulary. To realize the proposed approach, we first analyse schema.org to see how it can be placed in the well established semantic web services literature. Then we propose some minor extensions for necessary points. Afterwards, we present our mapping and wrapping approach for semantic lifting of the Web APIs in JSON-LD format and grounding of JSON-LD requests annotated with schema.org to the accepted data format of individual Web APIs.

The remainder of this paper is structured as follows: Section \ref{sec:relatedwork} presents a literature review on the semantic web services field with a focus on approaches for machine readable description of Web APIs. Section \ref{sec:actions} gives an introduction to schema.org actions and analyses it in terms of lightweight semantic web services. Section \ref{sec:methodology} explains our methodology for the mapping and implementation of a wrapper for existing Web APIs. Section \ref{sec:usecase} demonstrates the publication and consumption of such annotated Web APIs through our wrapper. Finally, Section \ref{sec:conclusion} summarizes the work and presents some concluding remarks and pointers to future work.

\section{Related Work}
\label{sec:relatedwork}
Machine readable description of web has crucial importance for tasks like partial generation of client code or automation of web service tasks. For the latter, not only machine readability, also the semantics of the functionality and information model model as well as the non-functional properties should be described. There are several efforts in the literature targeting SOAP services either in a bottom-up manner, where the semantic annotations are attached to WSDL files and top-down approaches where the services semantically described first and then grounded through WSDL.

One of the prominent works in this field is OWL-S \cite{Martin2007}. It uses OWL (Web Ontology Language) as a base for a semantic web service ontology. It provides mechanisms for describing functionality, behaviour and non-functional properties of a web service with Description Logic (DL) It has, however, no decoupled conceptual model and DL is not suitable for describing processes.
SWSF \cite{Battle}  is built on top of the experience gained from OWL-S. It aims to create a more expressive framework by using First Order Logic (FOL) instead of DL for process modelling, which is undecidable. Additionally, SWSF uses an extended version of the Process Specification Language for defining the behavioural aspects of the web service.
Web Service Modelling Framework (WSMF) \cite{Fensel2002} offers a full-fledged  decoupled approach for automating the whole lifecycle of web service consumption. It consists of a conceptual model, a modelling ontology WSMO, a structured set of languages WSML and an execution environment WSMX \cite{Roman2006}. Additional to the mechanisms to describe functional, behavioural and non-functional aspects of web services, unlike the other approaches, it provides a mechanism for mediation and goals which are distinguished from web service's functional description.

For the RESTful web services, the initial efforts were adapted from SOAP services mainly to ensure the interoperability.  In fact, the WSMO-Lite ontology \cite{vitvar_wsmo-lite_2008} was developed as a lightweight version of WSMO. It is more interoperable with W3C recommended technologies and is used as underlying semantic model for MicroWSMO , a language that extends hRESTS microformat for semantic description of RESTful services \cite{Roman2015}. 

Recently, several approaches specific to RESTful services have been proposed. The advantage of these approaches come with being relatively simple, because they do not carry mechanisms targeting SOAP web services. Since these approaches are completely REST oriented, they facilitate the creation of hypermedia-driven APIs. These APIs allow clients to be as generic (i.e. decoupled from the API) as possible and use the APIs with minimal apriori knowledge. This is the main difference between these approaches and API description languages like WADL\footnote{Web Application Description Language}, Swagger\footnote{https://swagger.io}, RAML\footnote{https://raml.org} and API Blueprint\footnote{https://apiblueprint.org}, since they merely give a structure to the documentation without any semantics and hypermedia-driven navigation possibilities. An active effort towards hypermedia-driven Web APIs is Hydra \cite{Lanthaler2013}. Hydra is a vocabulary for bringing RESTful APIs and linked data together without any well-defined complex semantics to lower the entry barrier and ease the adoption. It allows Web API owners to describe a machine-oriented API documentation which contains all accessible resources and a client can navigate the API using links and supported operations. There is also an active effort to align Hydra with schema.org actions (e.g. mapping of HTTP operations to schema.org) and the working group that maintains the vocabulary already contributed to the inclusion of actions in schema.org .
There are other efforts towards the same direction with some differences. RESTDoc \cite{John2013} has a similar principle as Hydra with certain implications about the underlying semantics (i.e. RDF(S)) for discovery and composition tasks.  The main drawback of this approach is that it is tied to a specialized microformat syntax.  Additionally, it is not clear how the behavioural aspects can be represented. Another approach is RESTDesc \cite{verborgh_functional_2012}, which uses N3-Notation syntax and semantics to represent preconditions and postconditions  for operations defined on the resources of a Web API. This gives a relatively well-defined semantics and reasoning support for automated discovery and composition. 

To the best of our knowledge, there has not been an effort so far to place schema.org actions into the semantic web services field except the alignment with the Hydra vocabulary. Incorporation of our approach and Hydra would also be an interesting research topic. Our approach is completely based on schema.org and its extensions to describe resources and their specific instances (request and responses). The actions can be defined standalone as well as attached to a specific instance. This would allow us to specify operations not only on resource level but also on instance level (e.g. number of children may be required for one type of room and optional for another in a hotel booking API). An interesting feature of schema.org actions vocabulary is that it provides a mechanism to attach high-level operations beyond HTTP CRUD operations. This provides an opportunity for (semi-)automatic generation of agents like dialogue systems \cite{Simsek2018}.  The schema.org vocabulary is already a de-facto standard for annotation of data on the web. The possibility of adding actions to existing annotations on web pages may have interesting implications such as "actionable knowledge graphs", where users not only query a graph, but also complete certain tasks in an automated fashion. Additionally, unlike other approaches, we propose an extension for schema.org actions to facilitate description of authentication mechanisms.

\section{Schema.org Actions as a Web API Annotation Vocabulary}
\label{sec:actions}
In this section we will give a brief introduction to the actions in the schema.org vocabulary. Then we will investigate the vocabulary and its semantics in the scope of lightweight semantic web services.

\subsection{Schema.org Actions}

The schema.org actions have been included to the core vocabulary in 2014 to give a mechanism to annotate not only static entities, but also actions that can be taken on them\footnote{http://blog.schema.org/2014/04/announcing-schemaorg-actions.html}. The action vocabulary is built around the schema:Action\footnote{Throughout the paper, the "schema" prefix will be used for http://schema.org/ namespace.}, which is the most generic type of action. In the context of schema.org vocabulary, the definition of an action is quite generic. In fact, alongside defining actions for operations that can be carried over HTTP, actions can be also used for describing links to mobile applications or even real word actions like "eating a pie". In the context of our work, naturally we only consider the actions that can be carried typically over a web resource.

The action annotations can exist in two different forms: (a) as stand alone annotations with a value in the schema:object property (b) as potential actions defined on the instances of any type. In the former case, an action can be defined as the root type of the annotation and entities are attached to the schema:object property. In the latter case, actions can be connected to the entities with the schema:potentialAction property. 

The invocation method of an action is defined with an entry point (schema:EntryPoint instance) which allows to specify an endpoint for sending a request and HTTP method to make the invocation. An entry point can be connected to an action with the schema:target property.
An action may have input parameters expected for invocation and output messages that it promises to return. These aspects are typically described with special property specifications such as \textless property name\textgreater -input and \textless property name\textgreater -output. These specifications can be defined on any action and the property values can be constrained, for instance, with a value range or whether the parameter is required.

\begin{figure}
\centering
\includegraphics[width=0.9\textwidth]{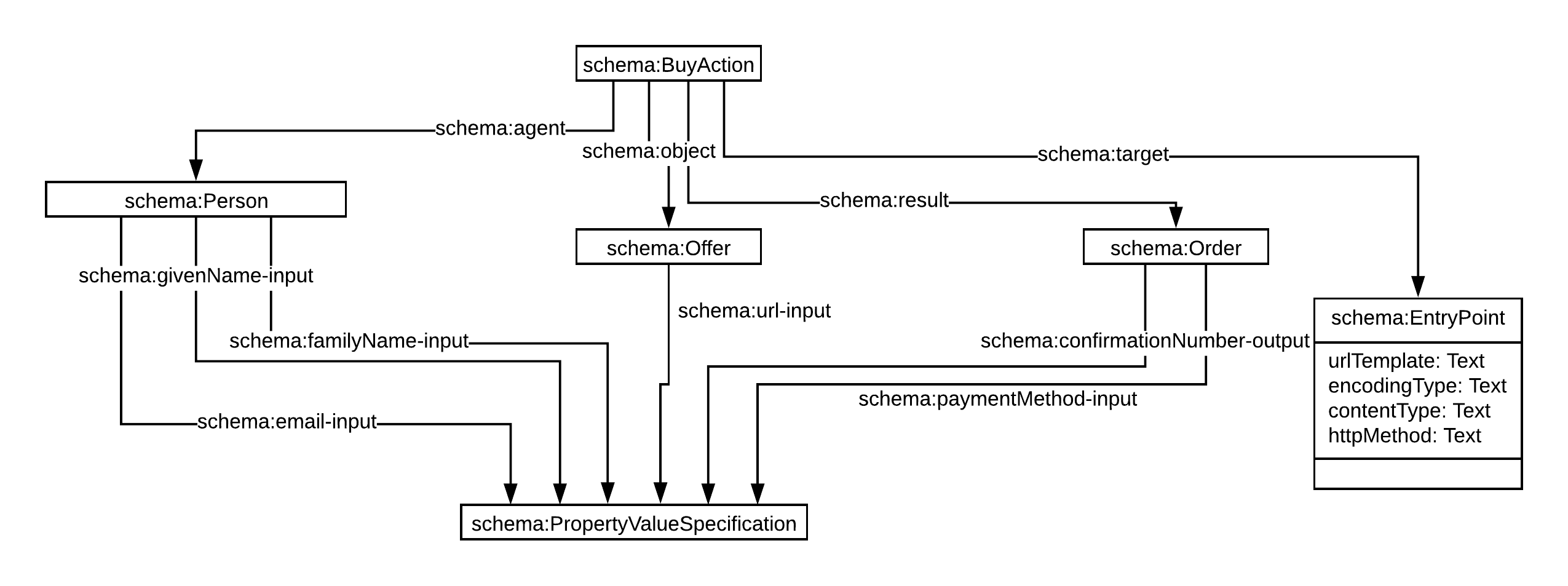}
\caption{Graphical representation of an example action}
\label{fig:action_example}

\end{figure}

Figure \ref{fig:action_example} depicts a schema:BuyAction with an schema:Offer attached as object. The action describes the entry point, which contains the target url for invocation alongside with encoding and HTTP method. Additionally, it defines that the result of this action invocation will be an instance of schema:Order type with a value for the schema:confirmationNumber property. The payment method, given name, family name and the email address of the agent who is carrying the action are requested to fulfil the action invocation. Note that, any property of any type attached to the action can be an input or output parameter regardless of being the object or the result of the action.

Schema.org does not provide any normative instructions for using the action vocabulary, only certain guidelines and suggestions\footnote{See https://schema.org/docs/actions.html}. They also do not adopt a strong formal semantics such as cardinality restrictions or range constraints to restrict certain actions on certain types. This provides us a room for refining and extending the vocabulary in order to create a lightweight semantic web service annotation vocabulary. In the next subsection we will discuss the schema.org actions vocabulary from the semantic web services point of view.

\subsection{Lightweight Web Service Description with Schema.org Actions}

Many Web APIs that operate on HTTP follows the REST principles \cite{Fielding2000} to some extent. That means, a Web API is a collection of resources ideally linked via certain operations. As it is defined in the REST architectural elements, \say{a resource is an identifiable abstraction of a set of entities.} The identification of resources are realized with URIs.  Whenever a resource is requested, the server maps the request to a set of entities and responses to the client with a certain resource representation and its metadata.

Table \ref{table:rest-sdo} shows a mapping between REST data elements and schema.org types and properties. Given the fact that most of the Web APIs are documented with HTML, a Web API can be seen as a set of schema:Action annotations on the documentation page(s).  The value of the schema:object property on a schema:Action instance represents the abstract concept of  a resource.  An action describes a high level operation on a resource (e.g. CommentAction, AddAction), which is in return connected to an HTTP method (e.g. POST, PUT). As for the parametrized operations, the input parameters can be described with  \textless property name \textgreater-input properties with schema:PropertyValueSpecification instances. The description of a parameter may include certain constraints inspired from HTTP forms (e.g. maximum length of a string, number intervals, whether the field is required).  Similarly, a resource can promise to return certain property values with  \textless property name \textgreater -output property definitions. Note that, the response to the resource request may have other property values as well, but the promised property values should definitely be included in the response.  The concrete representation of a request and response and metadata about how it should be interpreted by the client can be described with schema:encodingType and schema:contentType properties. 
Although action annotations are useful for describing the public interface of an API  to truly enable the hypermedia-driven nature, the concrete entities to which the resource requests mapped (i.e. responses) should be also described with their potential actions. This is achieved by attaching such actions to a Web API response with schema:potentialAction property.

\begin{table}
    \begin{tabular}{|l|l|}
    \hline
    REST Data Element                    & Relevant Schema.org Terms                              \\
    \hline
    Resource                             & schema:object Value \\
    Resource Method                      & schema:Action, schema:PropertyValueSpecification and schema:httpMethod                    \\
    Resource Representation and Metadata & schema:encodingType, schema:contentType                \\

\hline
    \end{tabular}
    \caption{Mapping of REST data elements to Schema.org}
    \label{table:rest-sdo}
\end{table}

\subsection{Service Semantics in Schema.org Actions}

In this section we analyse schema.org from a service semantics point of view. This analysis will be useful for the future, when the automation of web service tasks are considered. There is a distinction made between different type of service semantics in \cite{Sheth2003}, that is  adopted in a previous work \cite{Roman2015}, since they are important for automation of web service tasks like discovery/retrieval and composition. We analyse the schema.org actions vocabulary in the scope of the service semantics in the rest of this section. A summary of the analysis can be found in Table \ref{table:sdosemantics}.

\begin{table} [h]
    \begin{tabular}{|l|l|}
    \hline
    Type of Web Service Semantics & Schema.org Actions Support                                   \\
    \hline
    Information Semantics         & RDFS based Schema.org Vocabulary and its extensions          \\
    Functional Semantics          & Simple signature view based on input and output descriptions \\
    Behavioural Semantics         & Implicit through potential actions attached to responses     \\
    Non-Functional Semantics      & Limited support through the schema:instrument property       \\
    \hline
    \end{tabular}
    \caption{Summary of supported web service semantics by schema.org actions}
    \label{table:sdosemantics}
\end{table}

\subsubsection{Information Model and Semantics}

The information model defines the information transferred via the Web API, namely the inputs, outputs and fault messages.
An advantage in our approach is that we use a single domain ontology, namely schema.org,  and/or its extensions for the information model. Schema.org provides classes and properties to describe inputs and outputs. The property to attach fault messages to an action, schema:error, takes schema:Thing as range, which is the top concept in the vocabulary.
Schema.org has a very weak formal semantics, in fact, most of the semantics is embedded in the natural language descriptions of terms \cite{Schneider2014}. However, as the data model documentation\footnote{http://schema.org/docs/datamodel.html} states, its semantics is derived from RDFS.  It is relatively safe to apply RDFS entailment patterns regarding subclasses and subproperties (Table \ref{table:rdfs} \footnote{Adapted from Section 9.2.1 in https://www.w3.org/TR/rdf11-mt/}). However for domain and range definition schema.org introduces its own properties such as schema:domainIncludes and schema:rangeIncludes to avoid unintended inferences due to semantics of rdfs:range and rdfs:domain. Informally, the domain and range semantics are defined as follows:
\say{Each property may have one or more types as its domains. The property may be used for instances of any of these types.
Similarly, each property may have one or more types as its ranges. The value(s) of the property should be instances of at least one of these types.}
We make a closed world assumption and use domain and range definitions as constraints on the properties, meaning a validation would fail if a property has a value that is of a type not defined in the range or does not have a type at all. Similarly, it would fail, if a property is used on a subject that is not in the domain of the property or the instance has no type at all.
This closed world assumption is especially practical when it comes to narrow the range set of a property for an API in a certain domain. Several methods can be used for such a domain specification \cite{simsek2017domain} \cite{Knublauch}. 

\begin{table}
    \begin{tabular}{|l|l|p{3cm}|}
    \hline
    Contained Triple Pattern               & Added Triple Pattern    & Entailment Rule                      \\ \hline
    \textless A, rdf:type B \textgreater, \textless B, rdfs:subClassOf, C \textgreater & \textless A, rdf:type, C\textgreater        & Inheritance                          \\ \hline
    
    \textless A, rdfs:subClassOf, B\textgreater, \textless B, rdfs:subClassOf, C\textgreater                                      & \textless A, rdfs:subClassOf, C\textgreater                       & Transitivity of Subclass Relationship                           \\ \hline
 \textless A, rdf:type, rdfs:Class\textgreater              & \textless A, rdfs:subClassOf, A\textgreater & Reflexivity of Subclass Relationship \\ \hline
   \textless A, rdfs:subPropertyOf, B\textgreater, \textless B, rdfs:subPropertyOf, C\textgreater & \textless A, rdfs:subPropertyOf, C\textgreater & Transitivity of Subproperty Relationship \\ \hline
  \textless P, rdf:type, rdf:Property\textgreater              & \textless P, rdfs:subPropertyOf, P\textgreater & Reflexivity of Subproperty Relationship \\ \hline
    \textless A, rdfs:subPropertyOf, B\textgreater, \textless X, A, Y\textgreater & \textless X, B, Y\textgreater & Property Inheritance \\ \hline
    \end{tabular}
    \caption{RDFS semantics supported by Information Model based on Schema.org}
    
    \label{table:rdfs}
\end{table}

\subsubsection{Functional Description and Semantics}

The functional description specifies the functionality of a service, in other words, what a service can offer. A Web API described with schema.org is a collection of annotation in schema:Action type or its subtypes. Therefore there is no unified view to define the functional aspects of the entire service\footnote{There is schema:WebAPI pending for inclusion in the core vocabulary. This type would eventually provide the unified service view for the action annotations.}. However, we adopt a simple signature view \cite{Keller2006}, where the inputs and outputs of an operation is defined but no explicit mapping from inputs to outputs are given. 

\subsubsection{Behavioural Model and Semantics}
The behavioural model defines the order of operations to consume the service functionality. As for the Web APIs described with schema.org, there is no explicit behavioural model, but the order of operations implicitly dictated by the potential actions attached to the responses to a resource request. 

\subsubsection{Non-Functional Description and Semantics}
\begin{figure}
\centering
\includegraphics[width=0.8\textheight]{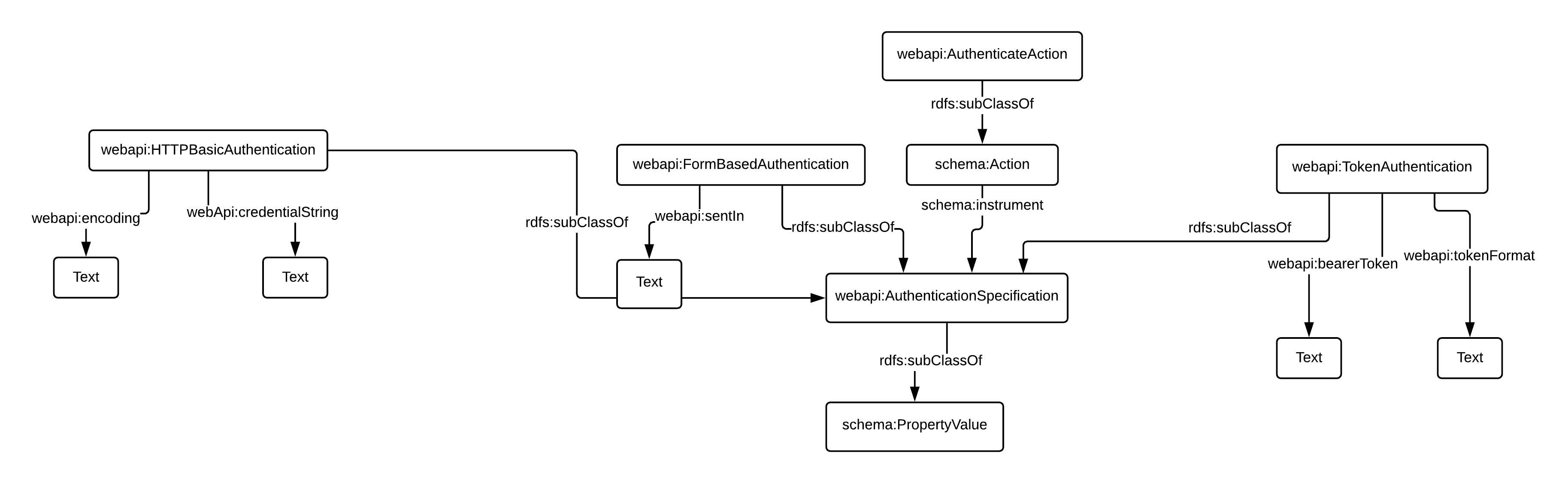}
\caption{Proposed schema.org action authentication extension}
\label{fig:sdo-auth}
\end{figure}

The non-functional description specifies the policies for the consumption of the service as well as the meta-information such as the creator and version of a given implementation of a service. Non-functional properties of SOAP web services were studied in previous work \cite{Toma2006}. These properties are including but not limited to temporal availability, price, payment and security \cite{o_sullivan_formal_2005}. As for the Web APIs, this aspect is often neglected or limited to meta-information about the service. Furthermore, the existing non-functional properties are usually described in a human readable way only, sometimes even outside of the service, in an API repository such as ProgrammableWeb\footnote{https://www.programmableweb.com}. In this work, we only focus on the security aspect due to lack of expressiveness of schema.org for such non-functional aspects. Schema.org offers a schema:instrument property which can be used to describe authentication tokens.
Figure \ref{fig:sdo-auth} shows the presented extension to describe authentication for the Web APIs\footnote{The proposed authentication extension is defined in https://actions.semantify.it/vocab/ and represented with 'webapi' prefix.}. 
 We define the authentication on the action, in other words at the resource method level. Based on the analysis we made on different API specification languages, identified three different authentication methods: token-based, basic and custom form-based authentication.

The token-based authentication is represented by webapi:TokenAuthentication type. The value of the bearerToken property on this type is mapped to the Authorization header in "Bearer \textless token\textgreater" pattern. Similarly, webapi:HTTPBasicAuthentication type is mapped to Authorization header in "Basic \textless token\textgreater" pattern. The third method is for the other custom authentication approaches. For these approaches, more information needs to be described such as, in which part of the request the token is sent (header, body or url) and what is the key and what is the value.  This can be done through using name and value properties of the schema:PropertyValue type which is the supertype of all authentication types. 

At the moment, we left the description of the workflow for obtaining the tokens open. This could be described by the API owner based on the protocol they use (e.g. OAuth 1.0/2.0). We define a webapi:AuthenticateAction type to facilate this description. The consuming clients should find these types of actions and follow the authentication workflow. In case the tokens are obtained outside of the API (e.g. manually through a user-interface), then the tokens can be directly used with individual actions of the Web API.

\section{Methodology for Wrapping Web APIs}
\label{sec:methodology}
To describe the methodology of wrapping let us first define the difference between mapping and wrapping. In our case mapping is the process of finding a vocabulary, certain types or classes and properties, and match them with a target entity. Here, the vocabulary is schema.org with certain classes and their properties. We try to represent the functionality as well as non-functional aspects of a Web API with the schema.org vocabulary, which means we map this API to schema.org. This can be done manually, semi-automatically or fully automatically. Wrapping, as opposed to this, is not a single mapping but a set of mappings which are applied to Web APIs. Wrapping is a repeatable process that is normally performed by a software, a wrapper. So first we define the mappings for the API and then, based on the mappings, a wrapper will be defined and implemented. In this section we describe the four-step process of defining mappings and a wrapper (Section \ref{sec:mapping}) and then the approach of simplifying the process by adding some degree of automation (Section \ref{sec:automating}).

\subsection{Mapping} 
\label{sec:mapping}
To find a mapping for an API, the first thing to do is the \textbf{API analysis}. We have to investigate what resource is the API serving, what HTTP method is this API accepting, what are the input parameters needed to fulfil the API's requirements, is there authentication necessary and of course what is the response object the API returns on a successful query.

Another important aspect to know about the API is whether it only serves one task (one-step API), where the work flow ends after one operation is fulfilled, or if it works on several tasks consecutively (multi-step or handshake API). Like in a checkout process, by multi-step or handshake we mean tasks where responses contain other possible tasks which can be triggered by an agent and has as well to be annotated. A simple example for a one-step task would be a weather report service where the input is a location and a time and the output is a weather report. As opposed to that a multi-step (or handshake) task would for example be a shopping service. The input to the first service, a search endpoint, is for example the size and the colour of  a t-shirt. The response is a list of results, where each result has a \textit{buy-action} attached. When executing the buy service the response would maybe be a bill with a pay action attached where the final response then is a payment confirmation.

The second step to find a mapping is the \textbf{vocabulary identification} to describe web services. In our case this vocabulary is schema.org with schema.org actions. Within that vocabulary for all input and output objects, the matching types have to be identified. And finally, within the types, the parameters that match the input and the output of the API have to be specified.

The third step is the \textbf{mapping implementation}. There are two different kind of mappings: (a) mapping of resources and (b) mapping of instances. The resource mapping can be seen as annotating the HTML documentation of an API. In Table \ref{tab:mapping}, a partial mapping of the \textit{/events/search/} resource of Eventbrite API\footnote{https://www.eventbrite.nl/developer/v3/endpoints/events/}. This mapping is then used to produce the action annotation in Figure \ref{fig:eventaction}. For the mapping of instances, the potential actions should be attached on the fly depending on the behaviour of the Web API. For example, the response to a request for the annotated resource would be an array of schema:Event instances and each of these instances could have an schema:AddAction as potential action with a schema:Ticket object.  

\begin{table}[ht]

\noindent\begin{minipage}{.49\textwidth}
\begin{tabular}{ | c | c |}
 \hline
 \multicolumn{2}{|c|}{\textbf{GET /events/search/ $\rightarrow$ schema:SearchAction}}  \\ \hline
 \multicolumn{2}{|c|}{\textbf{schema:object $\rightarrow$ schema:Event}}  \\ \hline
 \multicolumn{2}{|c|}{\textbf{schema:result $\rightarrow$ schema:Event}}  \\ \hline
 
 \textbf{Parameter name} & \textbf{Mapped Schema.org property} \\ \hline
  q & query \\ \hline
  location.latitude & location.Place.geo.latitude \\ \hline
  location.longitude &  location.Place.geo.longitude\\ \hline
  organizer.id & organizer.Organization.identifier \\ \hline
price & isAccesibleForFree \\ \hline
 \end{tabular}

\caption{Partial mapping of an Eventbrite resource to schema.org}
\label{tab:mapping}
\end{minipage}\hfill
\begin{minipage}{.49\textwidth}

\includegraphics[width=0.9\linewidth]{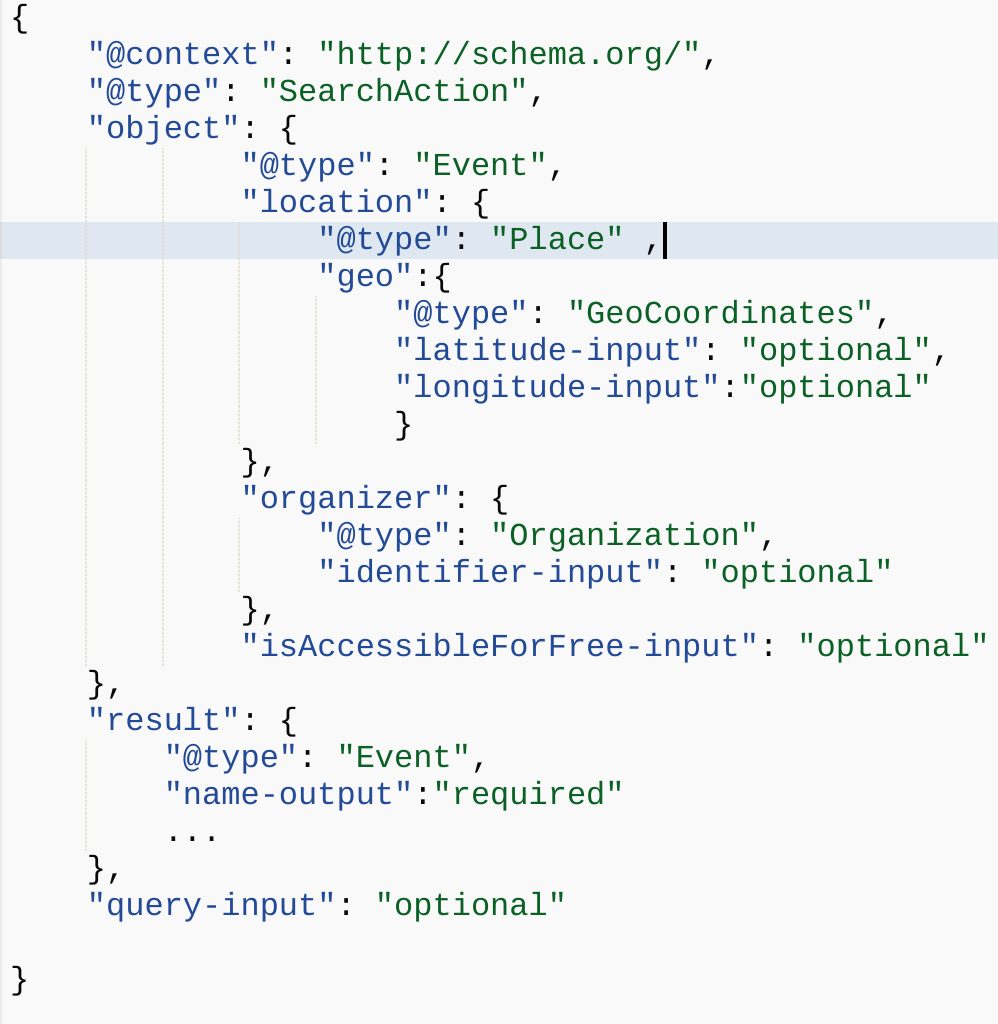}


\captionof{figure}{Partial JSON-LD representation of the Eventbrite resource in Table 4}
\label{fig:eventaction}
\end{minipage}\hfill

\end{table}

The forth and final step is the \textbf{wrapper definition/implementation}. The wrapper is mostly a standalone software which joins all mappings inside one application and which is flexibly designed for the use under various circumstances. If, for example, a booking API should be mapped, there will be mappers for the search service, the availability request, the booking, the payment and others. A wrapper joins these mappings inside a software and provides a starting point for agents using that software. The starting point itself is a schema.org annotated web service and here it makes sense to have a static annotation in the for of a JSON-LD snippet. From there the wrapper redirects all incoming requests in schema.org to the mapped API endpoint. The responses the endpoint replies with are taken by the wrapper, mapped to schema.org with the corresponding mapper and redirected to the agent who was initially requesting the resource. So the wrapper can be seen as a broker between a web service, an API, and an agent that understands schema.org. 

Following these four steps, under the assumption that certain implementation details are known, should be sufficient for manually building a working wrapper software that establishes communication between the annotated Web APIs and an agent. But the interesting question occurs if there is any potential for automation in any of these steps. The next section will briefly answer this question.

\subsection{Automating the mapping process} 
\label{sec:automating}
The advantages of manually mapping an API to a certain vocabulary are obvious. The flexibility of the concept selection process, the precision of the property selection the level of detail that can be respected and the 100\% certainty that the semantics of the API are represented the right way. On the other hand side, the disadvantages are equally obvious: manual mapping is a tedious process and does not scale. So the question arises, how to speed up this process by adding a certain level of automation.

We start by identifying the tasks where automation can be - partly or fully - applied, based on the four steps mentioned above. Provided that an API is described with a description language (e.g. Swagger), we can extract certain information from that descriptions. For the API analysis we can extract information like HTTP method. In general, the terms in an API description language can be mapped to schema.org types and properties until certain level, but the information model (e.g. schema.org types corresponding and input parameter) should be mapped manually since such description languages do not suggest any semantics. 

\section{Use Case}
\label{sec:usecase}
As a use case, or a proof-of-concept, we selected two similar APIs from the tourism domain. Tourism is a very convenient area for such a use case. There are a lot of websites with rich Web APIs for booking of products and services in tourism. Also schema.org is already widely distributed amongst hotels, as we could find out in two studies \cite{karle2016there,balcianalysis} and comprehensively applied to some destination management organisation (DMO) websites\cite{akbar2017complete}. In recent works we also worked on the extension of the schema.org vocabulary for the hotel domain\cite{karle2017extending}, so we know that vertical and its key figures very well.

The use cases focus on two aspect, namely the publication of actions as described above, and the feasibility of consumption of those actions by third party software. The implementation of the mappings was published on the \textit{semantify.it} platform \cite{karle2017semantify} as part of the \textit{semantify actions} wrapper.

The two use cases differ from each other on some aspects. Easybooking offers no real public API and the actions that were mapped had been detected by investigating the communication between the website and the server. Feratel provides a SOAP API. What both APIs have in common is their capabilities. They both provide search capability for hotels and booking functionality for hotel rooms, so the schema.org equivalents the implementation uses are schema:SearchAction and schema:BuyAction.

\subsection{Easybooking and Feratel API Mapping and Publication}
The web functionality we identified in the Easybooking use case are searching for a hotel room and booking a certain hotel room for a certain period of time. The search functionality was mapped to schema.org with schema:SearchAction. The search over the Easybooking API requires different input parameters which we mapped to schema.org with the properties schema:checkinTime, schema:checkoutTime, schema:numAdults and schema:numChildren. Note that these properties are not properties of schema:Action but of schema:LodgingBusiness and schema:HotelRoom. In this case schema:LodgingBusiness that contains a schema:HotelRoom is the value of schema:object property on schema:SearchAction annotation. If the search is successful, the response of the API is also mapped to schema.org. Here we define a multi-typed entity of schema:Offer and schema:LodgingReservation. Besides the obvious parameters we introduce an schema:itemOffered property to attach the hotel room that was searched for. The hotel room has some properties like schema:name and schema:description, as well as a schema:potentialAction parameter to define the next endpoint to contact, as the next step in the handshake API. The target endpoint of that action is the booking or "buy" endpoint which takes the schema.org multi-typed entity schema:Offer and schema:LodgingReservation as an input and responds with the final reservation confirmation in the form of a schema:LodgingReservation. The reason that the input object has to be a schema:LodgingReservation too is, that some properties required for booking are only available in the schema:LodgingReservation class. 

As for the Feratel API, we reused the same mapping. There was only one additional step involved to map XML and JSON-LD, due to Feratel's SOAP API.

\subsection{Consumption of Actions with a Dialogue System}
The actions in schema.org vocabulary allows us to define high-level operations on top of HTTP CRUD operations. For instance the semantics of reserving a room can be described with the schema:ReserveAction
or of buying an offer with the schema:BuyAction. This feature has also implications towards goal-oriented dialogue systems. The goal-oriented dialogue systems typically access a back-end system based on a user's intent to complete a task. The high-level resource-operation relationship can be converted to an intent (e.g. reserve.hotelroom intent based on schema:ReserveAction with a HotelRoom object). A Natural Language Understanding (NLU) module can classify user utterances to an extracted intent. The details of how the NLU trained for this intents is outside of the scope of this paper, but we demonstrate how a task-oriented dialogue system can consume Web APIs annotated with schema.org actions.

We implemented a dialogue system based on DialogFlow\footnote{https://dialogflow.com/} as NLU and programmatically added recognizable intents extracted from the action annotations from Easybooking and Feratel API \footnote{Demo dialog system: https://bot.dialogflow.com/3aa58719-b665-4e7b-970a-564c1b9a64c5}. The most of the existing industrial solutions require developing a dialogue system for each back-end, which is quite a tedious task. By extracting intents from annotated APIs, we simplify this task to a certain extent.

Listing \ref{listing:dialogueexample} shows an example dialogue generated from Easybooking API. First, Dialogflow classifies the user utterance to an intent extracted from the API. Then, the values for required slots (e.g. check-in and check-out time) for the intent are elicited from the user. The slots are also extracted dynamically from the functional description of the API. The dialogue continues until the intent is fulfilled.

\begin{lstlisting}[frame=single, label=listing:dialogueexample, caption=An example dialogue based for hotel booking over annotated Web APIs]
User: I am searching for the Hotel STIInnsbruck.
Bot: One item has been found. Hotel STIInnsbruck. You can ask me something about it.
User: Tell me more.
Bot:  You can search and find hotel rooms and offers.
User: Let's do it.
Bot: Please tell me your check-in time.
User: 1.1.18
Bot: Please tell me your check-out time.
User: 2.1.18
....
Bot: I found 4 items. The first 3 are: 1. Einzelzimmer, 2. Doppelzimmer, 3. Doppelzimmer Superior. 
User: 2
Bot: You can buy an offer of this room.
User: Let's do that.
...
Bot: Buy action completed. You can see the confirmation.



\end{lstlisting}

As demonstrated, action annotations  can ease the task related dialogue generation process by extracting intents and required slots. It should be also noted that the flow of the dialogue is determined on the fly based on the potential actions attached to responses of the API. For the dialogue in Listing \ref{listing:dialogueexample}, User utterance (Line 1) is first matched with a schema:SearchAction. The response to this action itself has a schema:SearchAction attached to it, which allows the user to search for hotel rooms and offers, after eliciting certain filtering criteria extracted from the action description (Line 2-11). The returned offers of hotel rooms also have schema:BuyAction as potential action, so the flow is completed after the buying process (Line 12-15). Note that the dialogue system presents several options to the user based on the potential actions attached to instances (Line 4 and Line 13).

\subsection{Other consumption} 
The Easybooking and the Feratel use case, implemented as part of the semantify.it actions wrapper, are used as part of the "publication heuristics" implementation\cite{kaerle2018heuristics}. The work's intention is to find heuristics for publishing dynamic, fast changing data, like for example availabilities of hotel rooms, as schema.org data. The problem is, that, if published in bulk, it would soon exceed network and server capabilities. So the authors found different publication heuristics where certain abstractions of the data are published and schema.org/Actions are attached to those objects. Those actions are generated by the wrapper mentioned in this paper.

\section{Conclusion and Feature Work}
\label{sec:conclusion}
In this paper, we showed how schema.org and schema.org actions can be used as a vocabulary to annotate Web APIs. We first analysed the schema.org vocabulary in terms of lightweight semantic web services and identified different semantics that can be useful to automate web service tasks in the feature. Afterwards, we explained different scenarios for creating mappings and implementing a wrapper that does the lifting and grounding for individual APIs. Additionally, we also showed how APIs annotated with actions can be consumed by automated agents like dialogue systems.

An obvious limitation of our approach is that schema.org is not expressive enough for every aspect of every domain. For instance, there is no way to describe a range for starting date for an event. This is a trade of at the moment between expressiveness and simplicity, we are trying to solve such issues while not introducing too drastic extensions to schema.org. Another limitation of our approach is the task of mapping creation being very tedious. We envision a machine learning approach to at least recommend actions for resources to speed up the mapping process. The details of these approach is a topic for further work.

Ideally, the APIs themselves should be semantically annotated to eliminate the need for a wrapper described in this paper which would not scale in long-term. However, we still argue that it is a necessary transition step to demonstrate the power of semantically annotated Web APIs to the API owners. JSON-LD format is very suitable for this task since it is a bridge between RDF and Web APIs; on the one hand it is suitable for resource identification and linking, on the other it is fully compatible with JSON format which is well known by developers and has very good tool support. We think this is the main advantage of JSON-LD, as it is also pointed out in Hydra\cite{Lanthaler2013}.

We also showed that one advantage of schema.org actions is that they can be directly attached to instances as potential actions and they can co-exist with the existing annotations on the web-pages. This would also help converting "knowledge graphs" to "service graphs" that contain actionable data.

\section*{Acknowledgement}
The authors would like to thank the Online Communications working group (OC)\footnote{http://oc.sti2.at} for their active discussions and input during the OC meetings and the members of the semantify.it team (Richard Dvorsky, Thibault Gerrier, Roland Gritzer, Philipp H\"{a}usle, Omar Holzknecht and Dennis Sommer) for their valuable input and their implementation effort. 





\bibliographystyle{model3a-num-names}
\bibliography{references}

\clearpage

\end{document}